\documentstyle[eqsecnum,multicol,aps,psfig]{revtex}
\draft

\headheight=\baselineskip
\begin{document}

\bibliographystyle{prsty}

%%%%%%%%%%   define here your own macros %%%
\def\ca{Ca^{2+}}
\def\caaq{Ca^{2+}_{(aq)}}
\def\oh{OH^{-}}
\def\ohaq{OH^{-}_{(aq)}}
\def\si{H_2SiO_4^{2-}}
\def\siaq{H_2SiO_{4(aq)}^{2-}}

\title{
\vspace*{-0.5cm}\hfill {\tt cond-mat/9508xxx}
       \vspace*{0.5cm}
       \\
A reaction-diffusion model for the hydration/setting of
cement}

\author{F. Tzschichholz, H.J. Herrmann and H. Zanni}

\address{
Laboratoire de Physique et M\'ecanique des Milieux H\'et\'erog\`enes
(CNRS, URA 857),\\ \'Ecole Sup\'erieure de Physique
et de Chimie Industrielle de la Ville de Paris,\\
10 rue Vauquelin, 75231 Paris Cedex 05, France}

\date{\today}
\maketitle

\begin{abstract}
We propose a heterogeneous reaction-diffusion model for the hydration
and setting of cement. The model is based on diffusional ion transport
and on cement specific chemical dissolution/precipitation reactions under
spatial heterogeneous solid/liquid conditions. We simulate the
spatial and temporal evolution of precipitated micro structures
starting from initial random configurations of anhydrous cement particles.
Though the simulations have been performed for
two dimensional systems, we are able to reproduce
qualitatively basic features of the cement hydration problem.
The proposed model is also applicable to general
water/mineral systems.
\end{abstract}

%61.43.-j  Disordered solids
%64.75.+g  Solubility, segregation, and mixing
%          (see also 82.60.L Thermodynamics of solutions)
%81.30.Mh  Precipitation
%81.35.+k  Granular materials: aggregation characteristics (e.g.,
%          grain size, particle size distribution, porosity)
%82.70.-y  Disperse systems
%82.20.Hf  Mechanisms and product distribution
%

\newsavebox{\boxl}
\savebox{\boxl}{$\ell$}

\begin{multicols}{2}
\section{Introduction}

In the present paper we propose a
heterogeneous reaction-diffusion model for the hydration and
setting of cement. The proposed model is based on the
experimental observation
that cement hydration can be described by a
dissolution/precipitation mechanism \cite{b0,b3,b8,b16,b12}.

The elementary aspects of the cement hydration/setting process
on a mesoscopic length scale can be characterized as follows.
At the initial stage cement particles or powder
(in our case tricalcium silicate, $C_3S \equiv Ca_3SiO_5$)
are mixed well with the solvent (water).
Rapidly after mixing the {\em dissolution} reaction
of the cement particles starts. Its principal reaction
products are ions which are mobile and
may {\em diffuse} into the bulk of the solvent
(in our case $\ca$ , $\oh$ and $\si$ ions). At this stage the ion
concentrations in the bulk of the solvent are very low and, as
a consequence, one finds strong ion fluxes from the dissolution
front away into the solvent. However, for a given temperature and
pressure the ion concentrations can not take arbitrary high values.
Rather, the ion concentrations are bounded by finite
solubility products above which solid phases start to {\em precipitate}
from the solution. There are two associated precipitation reactions:
a) the precipitation of calcium hydro-silicate sometimes referred as
`cement gel'
($C_{1.5}SH_{2.5} \equiv (CaO)_{1.5}\,(SiO_2)\,(H_2O)_{2.5}$)
and b) the precipitation of calcium hydroxide or `Portlandite'
($CH \equiv Ca(OH)_2 $)\cite{b8}.
While the growth of the cement gel is the
basis for the whole cement binding process, the growth of Portlandite
mainly happens in order to compensate for the accumulation of
$\ca$ and $\oh$ ions in solution (see below).

The process of cement dissolution, ion transport,
and cement gel/Portlandite precipitation is usually referred as
`cement hydration'.
This process is to a high extent heterogeneous in the sense, that
starting from initially random cement particle positions
dissolution and precipitation reactions change the
physical (boundary) conditions for the ion transports themselves.
This is due to the fact
that the diffusion within the solid phases can be neglected
relative to the solvent one.

Coupled chemical dissolution/precipitation reactions
are well known since a long time in geology and geochemistry.
As such kind of processes are fundamental for water/mineral systems
some effort has been undertaken to modellize
such systems \cite{b5,b14}. The approaches differ methodologically
and focus on different physical and chemical aspects.
Ref.~\ref{b5} considers the ion transport problem
for a single, solubility controlled, dissolution/precipitation reaction
employing a one dimensional cellular automaton approach.
A chemically more detailed approach of water/mineral interactions
includes considerations on the influence of
nucleation, electro chemistry and temperature
on chemical reaction rates, however,
without considering a transport equation\cite{b14}.

Recently, a relatively simple stochastic cellular automaton model
for cement hydration/setting has been proposed \cite{b6}.
Therein it has been tacitly assumed that ion transport is not relevant
for the hydration process above a certain mesoscopic
length scale (`pixel size').
Furthermore, in this model hydrate `pixels' perform a random walk
until they touch another solid `pixel', where they then
stick \cite{b6}.
In contrast to this
we position our deterministic model into the `opposite' direction:
mass transport happens due to diffusion of ions, and the hydrates
are regarded as immobile.

In Sec.~\ref{model} we give a detailed description of the employed
model. We consider heterogeneity aspects in Sec.~\ref{model-hetero},
chemical aspects in Sec.~\ref{model-reactions} and transport aspects
in Sec.~\ref{model-transport}.
Sec.~\ref{results} contains first results.
We present some images of calculated cement micro
structures, which
comprehensively demonstrate basic features and capabilities of the
present approach.
We consider parametric plots for the mean ion
concentrations, known in cement literature as `kinetic path
approach' and compare our findings with experimental
results \cite{b1}.
Following this we investigate the variation of the maximum average silica
concentration in solution for various chemical reaction rate
constants.
Finally we present curves for the hydration advancement in
time for two different sets of reaction rate constants.
In Sec.~\ref{conclusions} we summarize and give some
ideas about future work.

\section{The Model}\label{model}
In the following we describe the proposed
reaction diffusion model for the hydration and setting of
cement in more detail.

We  will first discuss how to quantitatively describe
the heterogeneities and outline some general features of
the employed model.
Following this we will summarize the dissolution/precipitation
reactions and the employed reaction rate laws.
Finally some physical aspects of the ion transport in the solvent are
discussed. The later one represents the {\em physical coupling}
between the dissolution and precipitation reactions.

\subsection{Heterogeneity aspects}\label{model-hetero}
We consider a discrete reaction-diffusion model in space
and time. The physicochemical system can be regarded as
being composed of `sufficient' small volume elements $V=\Delta x^3$
with $\Delta x$ typically between $10^{-6}m$ and $10^{-4}m$.
The initial configuration is determined from a digitized micro graph
image as follows:
 the micro graph is immediately taken
after mixing at time $t_0$,
one can assume to find only the solid phase $C_3S$
and water. By measuring the occupied volumes $V_i^{(\alpha)}$
of the phase $\alpha$, ($\alpha = aq, C_3S$), in each cell $i$ using
an image processing system, one determines the initial distribution
of mole numbers $n_i^{(\alpha)}(t_0)$ according to
$n_i^{(\alpha)}(t_0)= V_i^{(\alpha)}(t_0)/v^{(\alpha)}$.
The $v^{(\alpha)}$ are known molecular volumes for room temperature,
see Table I.
In general, such a micro graph shows cement particles of
different sizes immersed in water.
Because each cell volume is completely filled with solid
and solvent phases, one can calculate the solvent volume
of cell $i$ {\em at any time step}
from the corresponding solid volume(s),

\begin{equation}\label{solvent-volume}
V_i^{(aq)}(t) = V - \sum_{\alpha\ne aq} n_i^{(\alpha)}(t)
v^{(\alpha)}.
\end{equation}

This is a very  equation, because ion concentrations are
usually calculated with respect to the actual solvent volume
$V_i^{(aq)}(t)$.
The physical constraints of Eq.(\ref{solvent-volume}) are that
the chemical reactions must be {\em sufficiently slow}
in comparison to the solvent flow. Furthermore, it is
tacitly assumed that the whole system is connected to an external water
reservoir in such a way that the solid phases do not hinder the
flow.
%(compare for this our remarks about the
%ion transport due to {\em solvent flow} given below).

In a given volume element
the dissolution/precipitation reactions will, in general, not
instantaneously go to completion. The solvent volume
in each cell will rather increase/decrease
continuously, defining a solvent distribution field,

\begin{equation}\label{porosity-field}
\varepsilon_i^{(aq)}(t)=V_i^{(aq)}(t)/V.
\end{equation}
The solid phase volume fractions $\varepsilon_i^{(\alpha)}(t)$ are
defined analogously.\\
We believe in fact that the volume fractions
$\varepsilon_i^{(\alpha)}(t)$, reflecting the
actual solvent/solid distribution in the system,
are sufficient to characterize most of
the aspects of the hydration process on a
macroscopic length scale.\\
Though we do not have information about these distributions
on a length scale smaller than $\Delta x$ we will assume
that the reactants are homogeneously distributed in each
volume element. The volume fractions may then be interpreted
as probabilities to find
at a random position within cell $i$
the reactant $\alpha$.

\subsection{Chemical aspects}\label{model-reactions}
In the following we will summarize the
considered dissolution/precipitation
reactions and the employed reaction rate laws.
During the course of the simulation
it may happen that the reaction flux,
as calculated from the kinetic equations, does exceed the available
amounts of chemical reactants in a given cell.
We therefore check at each time step and for each chemical reaction
if there exists a limiting reactant. If this happens, we define the
reaction flux through the amount of the available limiting reactant.

\subsubsection{$C_3S$ dissolution}
The $C_3S$ dissolution reaction is a spontaneous and exothermic
{\em surface reaction}
which happens at the $C_3S$-solvent interface(s) \cite{b8}.
It can be considered as irreversible,

\begin{equation}\label{dissolution-reaction}
C_3S_{(s)} + 3H_2O_{(\ell)}
\stackrel{k_1}{\rightharpoonup} 3\caaq +
4\ohaq + \siaq.
\end{equation}
Here $k_1$ denotes an appropriate surface reaction
constant in units of $mol\,m^{-2}\,s^{-1}$.\\
Chemical reactions must be considered as local, i.e.,
each volume element is acting as a small and independent
chemical reactor as long as no transport occurs. However,
for a pronounced dissolution reaction in a given cell
one has also to consider the possible reaction amounts
originating from the interfaces with its neighboring cells.
This can be done by allowing cement dissolution in cell $i$ through the
electro chemicalsolvent of that cell and through a fraction $\gamma$ of the
solvent of the neighboring cells $j$,

\begin{equation}\label{dissolution-kinetics}
\nu_1^{(\alpha)} \, d\xi_{i,1}(t)=
\nu_1^{(\alpha)}\, k_1\, \Delta t\, \Delta x^2\, \bigl(
p_{ii}
+\gamma \sum_{j=nn(i)}
(p_{ij} - p_{ii})\bigr).
\end{equation}

Here $\nu_1^{(\alpha)}$ denote the stoichiometric numbers of
species $\alpha$ in reaction (\ref{dissolution-reaction}), see
Table I.
The changes
in mole numbers of species $\alpha$ in cell $i$
in a time intervall $\Delta t$ due to this reaction are
denoted by $\nu_1^{(\alpha)} \, d\xi_{i,1}(t)$.
The sum on the right hand has to be taken over all
next nearest neighboring cells of cell $i$, as indicated by
$j=nn(i)$.
The first term $p_{ii}=\varepsilon_i^{(C_3S)}\varepsilon_i^{(aq)}$
in Eq.(\ref{dissolution-kinetics}) describes the
dissolution within cell $i$ ($\gamma = 0$) due to a `typical' reaction
interface. It can be understood
as a probability to find the two reactants,
$C_3S$ and water, in contact at an arbitrary
chosen point within cell $i$.
The `chemically active interface' between cell $i$ and $j$
is given by $p_{ij}\Delta x^2=
\Delta x^2 \varepsilon_i^{(C_3S)}\varepsilon_j^{(aq)}$.
The constant $\gamma$ controls the degree of dissolution due to
this interface(s). We have used a value $\gamma =1/8$ in our
simulations, yielding comparable reaction rates from inner
and outer cell surfaces.
The determination of the reaction interfaces
in terms of cement and water volume fractions is similar to
the degree of surface coverage used in
Langmuir absorption theory \cite{b11}.
However, the absorption of water
on the cement's surface and the desorption of ions
from this surface are not necessarily the rate determining steps.
In fact, the dissolution reaction (\ref{dissolution-reaction})
involves various physical and  subprocesses
\cite{b12}
which will not be considered here.
%Furthermore,
%as one can see from the definition
%of the reaction interfaces they are in general {\em not symmetric}
%with respect to the cell indices $i$ and $j$.

\subsubsection{$CSH$ dissolution/precipitation}
The precipitation of $C_{1.5}SH_{2.5}$ is an endothermic reaction
%(reaction enthalpy $+25\,kJ\,mol^{-1}$ \cite{ref1})
which can be considered as
reversible \cite{b8},

\begin{eqnarray}\label{csh-precipitation}
C_{1.5}SH_{2.5(s)} &\stackrel{k_2}{\rightleftharpoons} &\siaq \nonumber\\
&+& 1.5\caaq  + \ohaq +H_2O_{(\ell)},
\end{eqnarray}
having a forward (dissolution) rate constant $k_2$
(in units of $mol\,m^{-2}\,s^{-1}$).

We note, that the $CSH$ precipitation, i.e., the backward reaction
in Eq.(\ref{csh-precipitation}) is the main reaction in this
balance equation.
The reactants and products are considered to have a
fixed stoichiometry.
However, it is known from experimental
data that the stoichiometry of $C_xS_yH_z$
can be rather
variable (`solid solution') \cite{b3}.
We will not include these complications
into the model. Instead we consider a fixed and typical
calcium/silica ratio equal to $1.5$.

In  equilibrium the $C_{1.5}SH_{2.5}$
dissolution/precipitation reactions are  controlled
by an empirical value of solubility, giving the maximum amount
of solid one can dissolve in aqueous solution at a given
temperature (room temperature) and pressure (normal pressure).
Employing this empirical solubility constant one can directly
determine the equilibrium solubility product,
$S_2^{equi}=S_{2i}\vert^{equi}=
(\ca)_i^{1.5}(\oh)_i(\si)_i\vert^{equi}
\approx 1.2\times 10^{-9}\,mol^{3.5}\,liter^{-3.5}$, comp. Table I.
The square brackets denote
the ion concentrations with respect to the
available solvent, i.e.,
$(\ca)_i=n_i^{(\ca)}/V_i^{(aq)}$ etc..
When in a given cell the ion product
is larger than the solubility product, locally precipitation happens.
If it is lower, $C_{1.5}SH_{2.5}$ becomes dissolved.
The $CSH$ dissolution is only marginal and for this reason we will
employ for the rate of dissolution a simpler expression than
for the $C_3S$ dissolution reaction.

We assume that the dissolution reaction in cell $i$ is proportional
to the typical
reaction interface $\Delta x^2 \varepsilon_i^{(aq)}$ with
the dissolution constant $k_2$ as the constant of
proportionality.
The precipitation reaction is assumed to be proportional to the
reaction interface and proportional
to the ion product $S_{2i}$, as defined above.
The rate constant of the precipitation reaction
can be eliminated, by use of the equilibrium condition, $d\xi_{i,2}=0$.
We have employed the following rate equation,
\begin{eqnarray}\label{precipitation-kinetics1}
\nu_2^{(\alpha)} \, d\xi_{i,2}(t) =
\nu_2^{(\alpha)}\, k_2\, \Delta t\, \Delta x^2\, \varepsilon_i^{(aq)}
\bigl(1- \frac{S_{2i}}{S_2^{equi}}\bigr).
\end{eqnarray}

The stoichiometric numbers again are denoted by $\nu_2^{(\alpha)}$,
see Table I, and the change in mole numbers of
species $\alpha$ due to this reaction is
$\nu_2^{(\alpha)} \, d\xi_{i,2}(t)$.

\subsubsection{CH dissolution/precipitation}
The precipitation of calcium hydroxide accompanies
the $CSH$ precipitation.
This is because the $CSH$ precipitation does not consume
the ions in the same proportions as they are released due to $C_3S$
dissolution.
The non-reacted $\ca$ and $\oh$ ions soon begin to
accumulate in solution, until the solubility
of $CH$ is exceeded. However, the precipitation of $CH$ and $CSH$
are, in general, not simultaneous because the corresponding
solubilities are very different, comp. Table I.
The dissolution/precipitation of $CH$
%(reaction enthalpy $+16\, kJ\, mol^{-1}$ \cite{ref1})
can be considered as reversible,

\begin{equation}\label{ch-precipitation}
CH_{(s)} \stackrel{k_3}{\rightleftharpoons} \caaq   + 2 \ohaq ,
\end{equation}
having a dissolution constant $k_3$ ($mol\,m^{-2}\,s^{-1}$).

The corresponding solubility product for $CH$ is defined as
$S_3^{equi}=S_{3i}\vert^{equi}=
(\ca)_i(\oh)_i^2\vert^{equi}
\approx 3.3\times 10^{-6}\, mol^3\, liter^{-3}$
which is about three orders of
magnitude larger than $S_2^{equi}$, see Table I,

\begin{equation}\label{precipitation-kinetics2}
\nu_3^{(\alpha)} \, d\xi_{i,3}(t) =
\nu_3^{(\alpha)}\, k_3\, \Delta t\, \Delta x^2\,
\varepsilon_i^{(aq)}
\bigl(1- \frac{S_{3i}}{S_3^{equi}}\bigr).
\end{equation}

In all other respects the $CH$ reaction is treated analogously to
the $CSH$ dissolution/precipitation reaction.

\subsubsection{Chemical shrinkage}

One can write down from Eqs.(\ref{dissolution-reaction}),
(\ref{csh-precipitation}) and (\ref{ch-precipitation}) the
net reaction for the cement hydration process,

\begin{equation}\label{net-reaction}
C_3S_{(s)} + 4H_2O_{(\ell)}
\rightharpoonup C_{1.5}SH_{2.5(s)} + 1.5  CH_{(s)}.
\end{equation}

Calculating for the above equation the occupied volumes for one mole
one finds for the left side approximately $144\, cm^3$ and for the
hydrate products $130\, cm^3$, comp. Table I.
Hence, the reaction products occupy
a volume around 10\% smaller than the reactants (including solvent).
This effect, which
is typical for hydraulic binders, is termed `chemical shrinkage' or
Le Chatelier effect. It is responsible for the transition of cement
suspensions or pastes having a finite viscosity and zero elastic
moduli to rigid hydrated cement having an infinite viscosity and
finite elastic moduli. Wherever topologically possible water flow
will try to compensate for the loss of volume. However, in the
course of an experimental cement setting and hardening (rigidification)
process available
solvent flow paths may vanish and volume loss becomes partially
compensated building up mechanical deformations (stresses)
within the solid phase agglomerate. In turn these stresses influence
the reaction rates (which in general are pressure dependent) in a
complex manner \cite{b16}.
The appearence of voids (pores) on the submicron scale
is experimentally also well established \cite{b0}.

We will not modellize such complex problems as mechanical stresses,
solvent cavitation and solvent flow
due to chemical shrinkage in our present approach.

Instead we will assume throughout all calculations that the solvent
is able to balance the chemical shrinkage for all time steps and
for all volume elements according to Eq.(\ref{solvent-volume}).
This assumption is equivalent to the
introduction of local source terms for water, i.e., the hydrating
system is connected to an external water reservoir in an
appropriate way.

We would like to point out that
Eq.(\ref{net-reaction}) is an overall reaction which per definition
only holds for an isolated system. On the other hand one
cannot treat the volume elements as chemically isolated,
simply because of the ion fluxes going through the elements.
This and the different values for the $CSH$ and $CH$ solubilities
imply in general locally non-congruent precipitation reactions.
As a consequence one cannot simply `replace' one
dissolved unit volume of $C_3S$
by 1.7 unit volumes $CSH$ and 0.6 unit volumes $CH$ as
has been proposed in \cite{b6}.

\subsection{Transport aspects}\label{model-transport}

It is known from experiments that ion transport due to
diffusion is fundamental
in cement chemistry, because of very strong ion concentration
gradients close to the dissolving $C_3S$-solvent interface(s).
In general one
should consider the convective diffusion equation, however,
this would imply to solve the
hydrodynamic equations.
In the present paper we consider the
usual diffusion equation with local, appropriately
defined transport coefficients. It is in the
context of the investigated problem useful to consider only the ion
diffusion within the solvent. For convenience we document in
Table I the diffusion constants $D^{(\alpha)}$
as calculated from
the electric mobilities at room temperature and normal pressure
for infinite dilution. The solids and the
solvent's diffusion constants are set to zero \cite{b13}.
This allows a relatively compact notation for the equations
of continuity,

\begin{eqnarray}\label{continuity-equation}
\Delta n_i^{(\alpha)}&=&
 \frac{\Delta t}{\Delta x^2}V
    \sum_{j=nn(i)}
\bigl(
D_{ij}^{(\alpha)}c_j^{(\alpha)}
-D_{ij}^{(\alpha)}c_i^{(\alpha)}
\bigr) \nonumber \\
&+& \sum_{k=1}^3 \nu_k^{(\alpha)}\, d\xi_{i,k}.
\end{eqnarray}

The second term on the right side of Eq.(\ref{continuity-equation})
represents the
already known source term due to the chemical reactions
Eqs.(\ref{dissolution-kinetics}), (\ref{precipitation-kinetics1})
and (\ref{precipitation-kinetics2}).
The $D_{ij}^{(\alpha)}$ are the transport coefficients, that depend
on the solvent volume fractions (porosities) of the cells involved
in the diffusion process,
$D_{ij}^{(\alpha)}=D^{(\alpha)}\varepsilon_i^{(aq)}\varepsilon_j^{(aq)}$.
The ion concentrations are taken with respect to the actual
solvent volume, i.e., $c_i^{(\alpha)}=n_i^{(\alpha)}/V_i^{(aq)}$.
We have employed periodic boundary conditions
in all calculations.

The general formulation of the ion transport problem
in cement chemistry
involves various subproblems which we will not consider at this
stage of modelization, as there are the transport due to

\subsubsection{Heat conductance}
The
cement dissolution reaction is strongly exothermic.
Analogous to the concentration gradients one finds near the
various dissolution fronts strong temperature gradients,
leading to  non isothermal solvent flow fields, with its complications.
The heat redistribution is probably also of direct
importance for the dissolution/precipitation
reactions, because their reaction constants
as well as the corresponding solubility products
usually depend very sensitively on the temperature. We will
neglect this possible effects in our approach assuming an
overall constant room temperature, $T_0=298$ Kelvin.

\subsubsection{Electrostatic interactions}
Because one is dealing with the transport of electrically
charged particles, one must consider, in general, Poisson's equation
for the electrostatic potential in order to determine the
resulting ion mobilities due to their electric field.
We will neglect in our simulations
all electrostatic interactions. Furthermore,
all chemical rate calculations are done
in terms of molar concentrations and not in terms of
`chemical activities' (which is equivalent in the limit
of infinite dilution). Hence we will treat the ions in solution
as uncharged particles throughout all calculations.

\subsubsection{Solvent flow}
 This is
possibly of importance for the overall hydration process,
because the solvent has to move during the
dissolution/precipitation reactions accordingly, in order to
compensate for the loss/gain of solid volume,
creating advective ion fluxes. For an infinite, plane cement-water
interface one might argue that the solvent flow is always
perpendicular to the interface. Under this particular conditions
it can be shown by scaling arguments that
advective ion transport is
unimportant relative to the diffusion transport, because of
the relatively low molecular volume of cement, comp. Table I.
However, in a finite heterogeneous system solvent flows tangential
with respect to the interfaces are also expected and the argument
given above is limited.
In principle one has to consider the equations of
hydrodynamics for this purpose. However, this moving boundary
problem is complex and we will not consider it in this form here.
Instead we will determine the spatial and temporal
solvent distribution $V_i^{(aq)}(t)$ according to
Eq.(\ref{solvent-volume}).

\section{Results}\label{results}
{}From a general point of view cement hydration can be regarded
as a heterogeneous (non equilibrium)
solid phase transformation forming from an anhydrous solid
phase two hydrated solid phases\cite{b17}.
In contrast to known solid-solid phase transformations for example
in metallic alloys, cement hydration bases strongly on the
presence of a solvent phase (mostly water).
The solvent controlls the transformation in a twofold sense: it is
directly part of the chemical reactions and on the other hand
it controlls the ion transport to a very large extent.

\begin{figure}[htb]
\centerline{\psfig{file=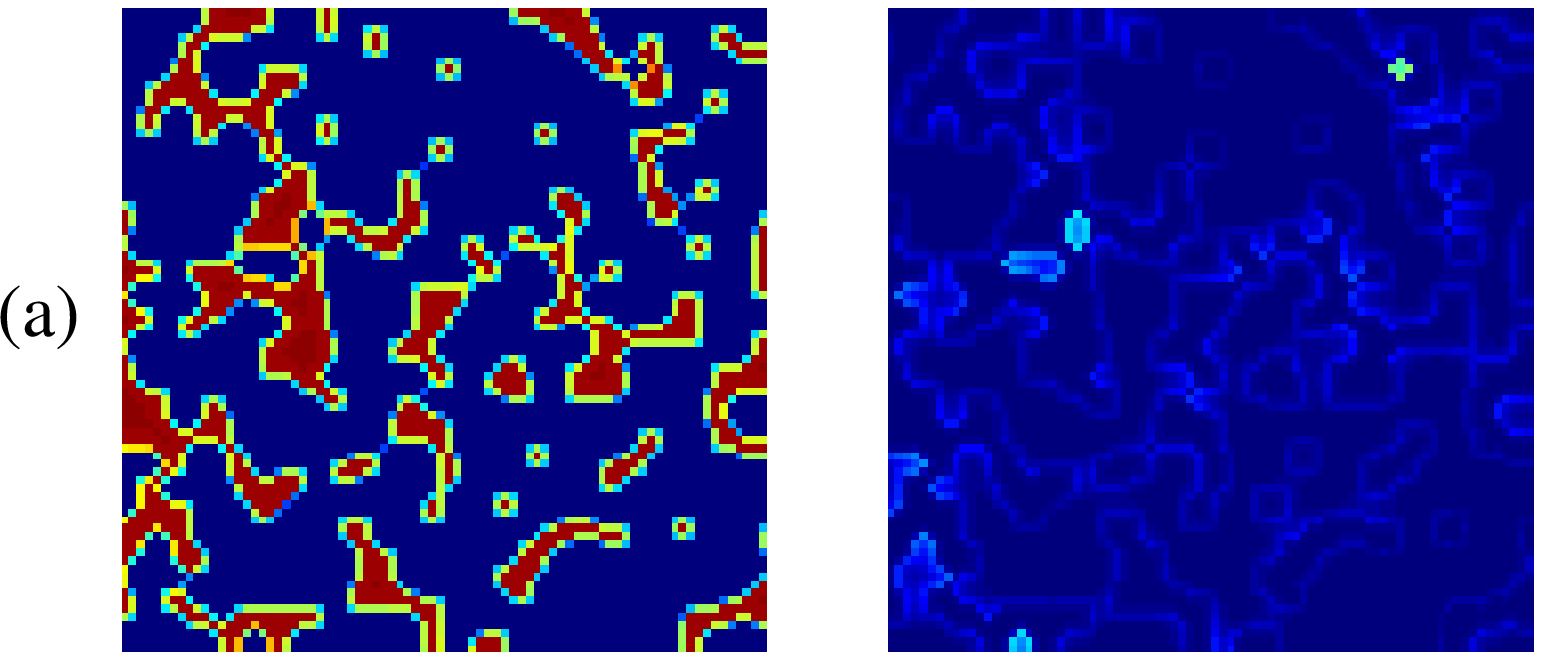,width=\textwidth}}
\vskip 5mm
\centerline{\psfig{file=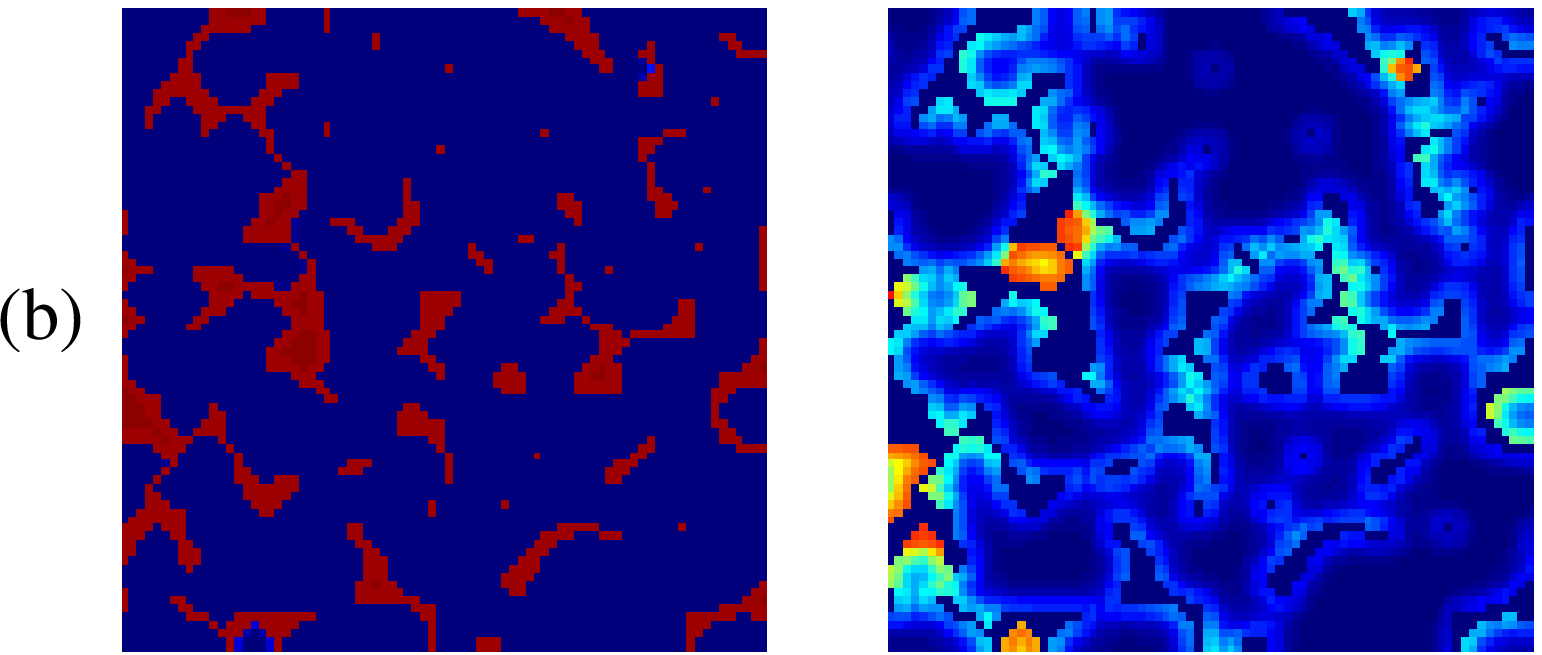,width=\textwidth}}
\vskip 5mm
\centerline{\psfig{file=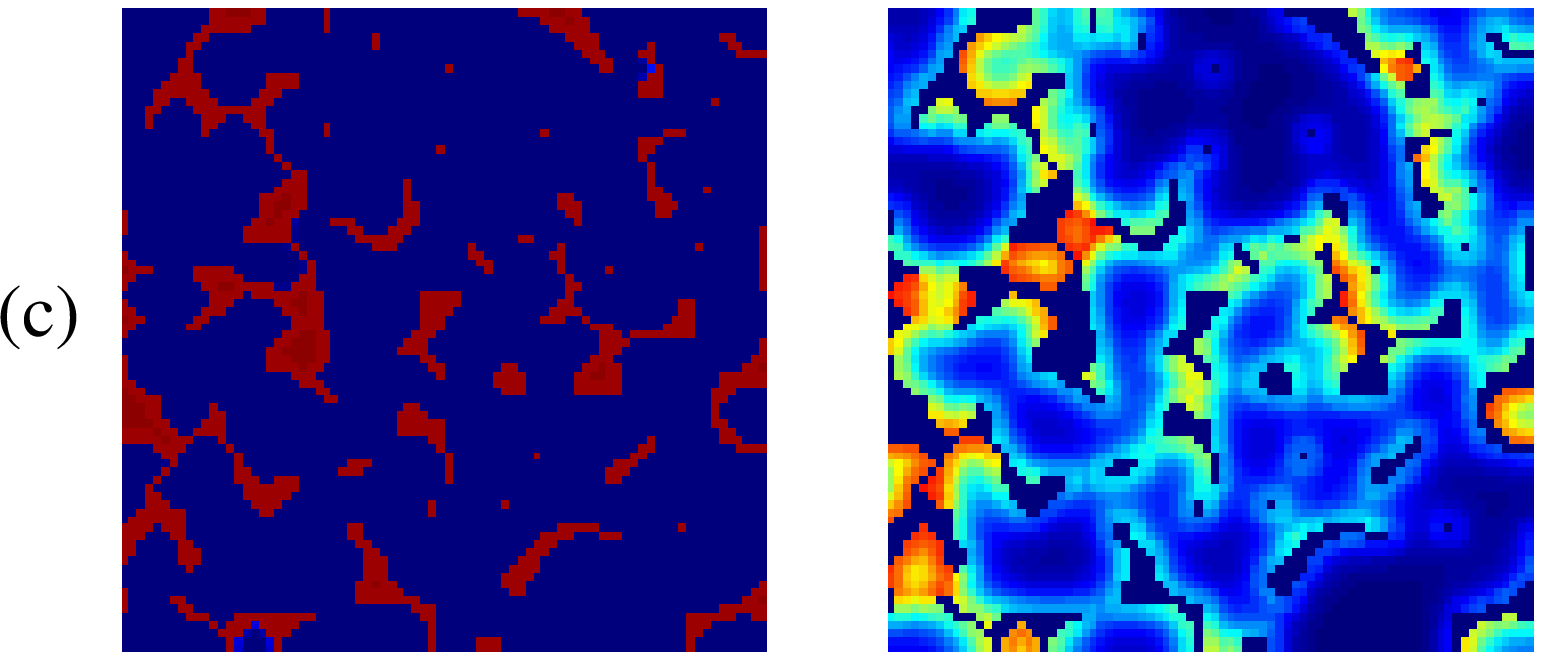,width=\textwidth}}
\vskip 5mm
\caption{
  The above figures show numerically calculated cement micro
  structures after (a) $t=1500\,s$, (b) $t=2500\,s$ and
  (c) $t=5000\,s$.
  Images on the left side show the
  unhydrated cement phase (C3S) while those on the right
  side represent the hydrated,
  precipitated cement (CSH). The colors indicate the local solid volume
  fractions, i.e., blue (0-20\%), cyan (20-40\%), green (40-60\%),
  yellow-orange (60-80\%) and red (80-100\%). The initial water/cement
  ratio is about 0.4, the reaction rate constants are $k_1=10^{-3}$,
  $k_2=10^{-12}$ and $k_3=10^{-10}$ (in units of
  $mol\,m^{-2}\,s^{-1}$).
  The employed cell size is $\Delta x=10^{-4}\,m$, the linear system size
  $L=100\Delta x$, and the linear cement particle size $\ell =5\Delta x$.
  The time step is $0.01\,s$. All other parameters as in Table I.
           }
\label{fig:images}
\end{figure}

The complexity of the problem and the features of the present
approach are most convenient illustrated by Fig.~\ref{fig:images}.
Therein we show the solid volume fractions of anhydrous
cement, $\varepsilon^{(C_3S)}$ (left side), and of hydrated cement,
$\varepsilon^{(CSH)}$ (right side), for three different hydration times
(a) $t=1500\,s$, (b) $t=2500\,s$ and (c) $t=5000\,s$. The occupied
volume fractions are color coded ranging from blue (0-20\%),
cyan (20-40\%), green (40-60\%) over yellow-orange (60-80\%) to
red (80-100\%).
The calculated micrographs Fig.~\ref{fig:images}a-c
show, as expected, spatially inhomogeneous
nucleation and growth of $CSH$ hydrate.
The hydrate precipitation and the associated growth
of hydrate surface layers surrounding dissolving
cement particles is relatively slow.
However, in Fig.~\ref{fig:images}b one
especially observes that the
rate of $CSH$ precipitation is spatially
strongly varying (notice the red and cyan
spots in the right image). This reflects the spatial fluctuations
in chemical reactivities (local $C_3S$ surface/volume ratio), i.e.,
the reactivity of a `fjord' is higher
than that of a big `lake'.
Supported by experimental observations it has been argued that
cement particle surfaces close to each other may indeed act as
very strong inhomogeneities leading to localized $CSH$ nucleation
forming `bridges' between adjacent particles \cite{b9}.
Our micrographs confirm this picture. We note, that the proposed
model does not include any (auto-)catalytic effect of $CSH$ on
its growth yet, compare Eq.~(\ref{precipitation-kinetics1}).
This point is intensively studied at the moment.

As the hydrate precipitation/dissolution reaction rates
depend on the local supersaturation values, the interesting question
arises how global characteristic points of the cement hydration
kinetics can be defined.
In a recent experimental work it has been proposed to characterize
different kinetic regimes occuring during the $C_3S$ hydration
by means of parametric plots of `typical'
calcium versus silicate concentrations as estimated from
calorimetric or conductimetric measurements
(`kinetic path approach')\cite{b1}. The experiments have been
conducted employing stirred diluted $C_3S$ suspensions having a
water/cement weight ratio between $10$ and $50$.
However, to obtain typical quantities of such diluted suspensions
employing numerical simulations one would have to consider very
large systems. Instead we will compare the experimental suspension
data with numerical cement paste data having a water/cement ratio
close to the practically important value $0.5$.

\begin{figure}[htb]
\centerline{\psfig{file=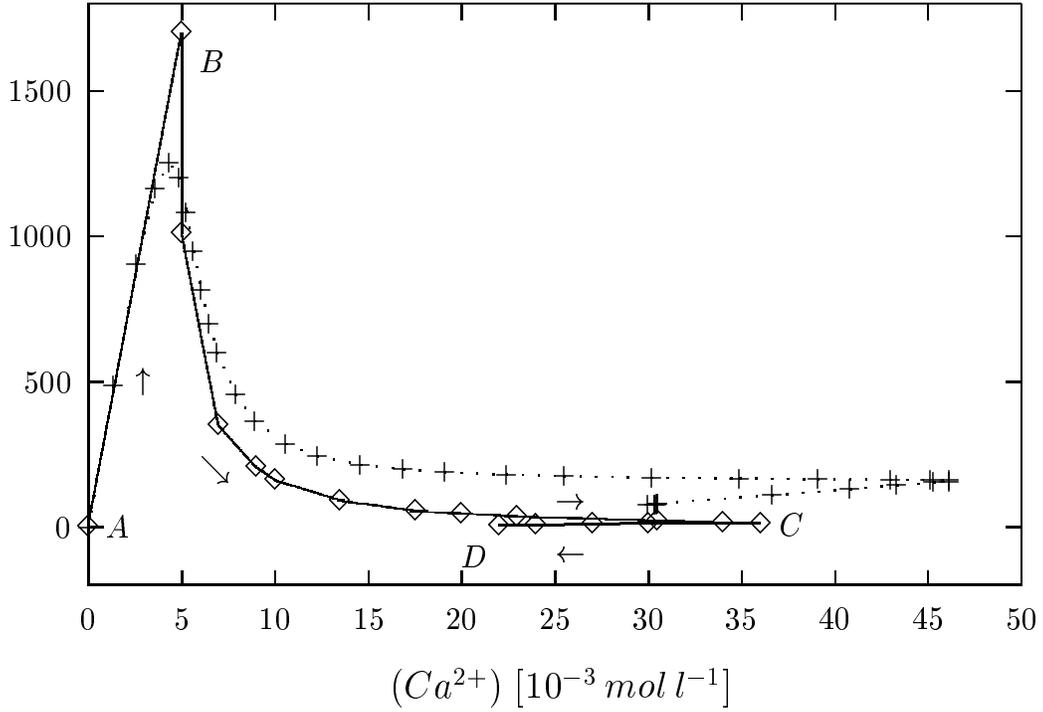,width=\textwidth}}
\caption{
  Parametric plot of the evolution of silica versus calcium
  concentration (`kinetic path approach'); $(\diamond)$ experimental
  results after Ref.~\ref{b1}, $(+)$ numerical results.
  The employed reaction rate constants are
  $k_1=5\cdot 10^{-5}$, $k_2=10^{-6}$ and $k_3=10^{-8}$ (in units of
  $mol\,m^{-2}\,s^{-1}$).
  The linear system
  size is $L=50\Delta x$, the cell size $\Delta x=10^{-4}\,m$ and
  the linear cement particle size $\ell =5\Delta x$.
  The employed time step is $0.1\,s$. Initial water/cement weight
  ratio $0.6$. All other parameters as in Table I.
          }
\label{fig:fig1}
\end{figure}

For convenience we reproduce in Fig.~\ref{fig:fig1}
some of the experimental data $(\diamond)$ of Ref.~\ref{b1}.
Initially one finds an `inductive'
cement dissolution period (period $AB$ in Fig.~\ref{fig:fig1}).
During this period the bulk solution is everywhere undersaturated
with respect to $CSH$ and $CH$, the concentration gradients
close to the cement particles are very high, and ions
rapidly distribute into the bulk solvent. We have
observed in our simulations that $CSH$ is already formed during
the `induction' period, however, to a very low extent. Because
of the initially very high concentration gradients
the ion products exceed the
$CSH$ solubility only in a very thin layer surrounding the cement
particles.
As the ions cummulate in solution the thickness of this
precipitating layer increases due to the broadening of the
ion distributions.

The precipitation counteracts
the further increase in silica concentration
both by chemical reaction and
by decreasing the cement particles surface permeability.
In result the silica concentration passes through a maximum
(point $B$ in Fig.~\ref{fig:fig1}), which might be interpreted as a
{\em global measure} for the onset of $CSH$ precipitation.
During the period $BC$ the silica concentration
decreases, while the calcium and hydroxyl concentrations continue
to increase as the $CSH$ precipitation consumes only a fraction of
these ions, see Eqs.~(\ref{dissolution-reaction})
and (\ref{csh-precipitation}).

\begin{figure}[htb]
\centerline{\psfig{file=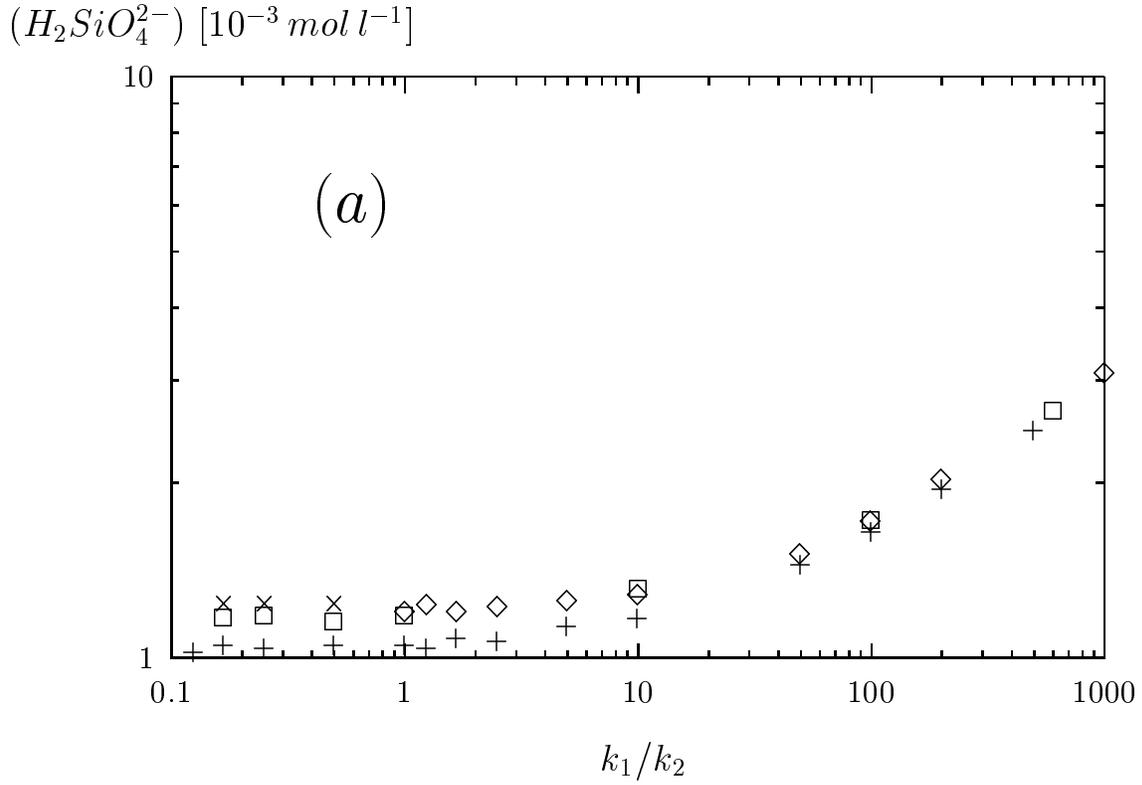,width=\textwidth}}
\centerline{\psfig{file=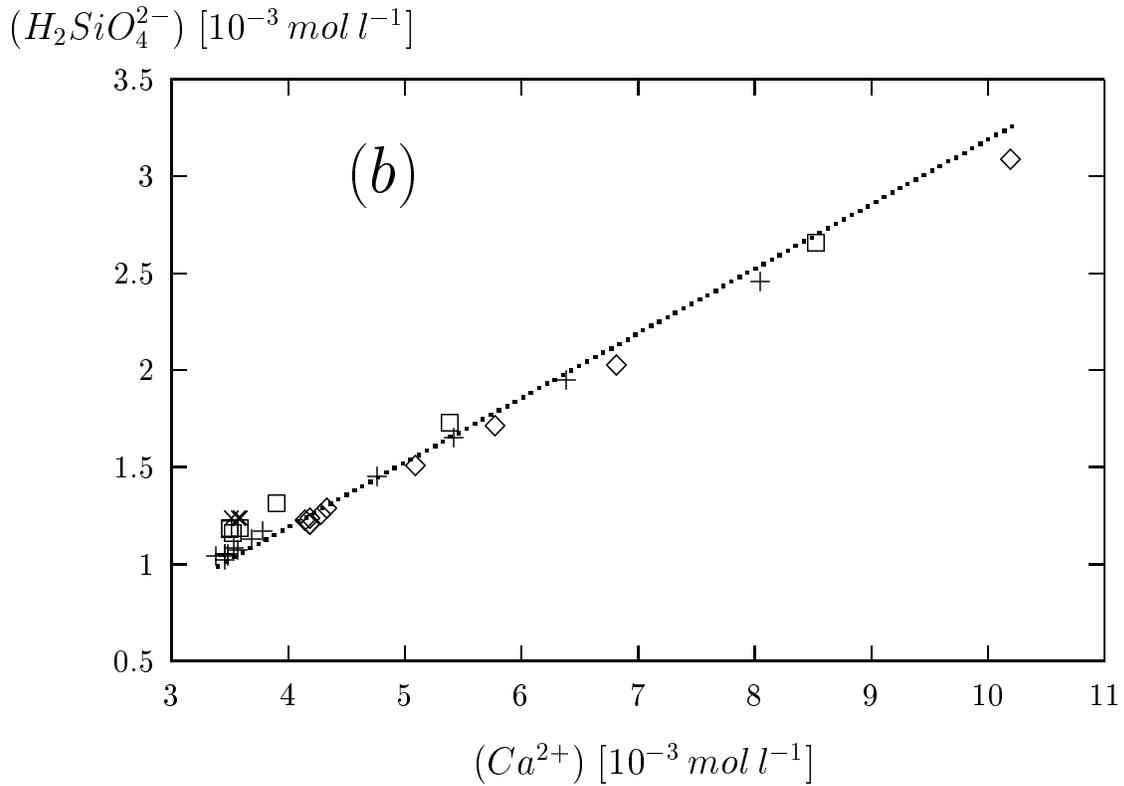,width=\textwidth}}
\caption{
  Calculated silica ion concentration $(H_2SiO_4^{2-})$ at the
  onset of $CSH$
  precipitation (corresponding to point $B$ in Fig.~\ref{fig:fig1}):
  (a) double logarithmic plot
  versus the relative reaction rate constant $k_1/k_2$, (b) plot
  against the momentaneous $(Ca^{2+})$ ion concentration.
  Symbols:
  $(\diamond)$ $k_1=10^{-4}\,mol\,m^{-2}\,s^{-1},\,k_1/k_2$
  ranging between $1$ and $1000$;
  $(+)$ $k_1=10^{-5}\,mol\,m^{-2}\,s^{-1},\,
  k_1/k_2$ ranging between $0.125$ and  $500$;
  $(\Box)$ $k_1=10^{-6}\,mol\,m^{-2}\,s^{-1},\,
  k_1/k_2$ ranging between $0.167$ and $600$;
  $(\times)$ $k_1=10^{-7}\,mol\,m^{-2}\,s^{-1},\,
  k_1/k_2$ ranging between $0.167$ and $0.5$.
  Each point corresponds to an average over $10$
  configurations having an initial water/cement
  weight ratio of about $0.5$.
  All other parameters as in Fig.~\ref{fig:fig1}.
          }
\label{fig:fig3}
\label{fig:fig2}
\end{figure}

Typically after a couple of hours
the solution becomes oversaturated with respect to calcium
hydroxide $CH$ (point $C$ in Fig.~\ref{fig:fig1}).
Both hydrates
precipitate very slowly
(experimentally several weeks between point $C$ and $D$),
lowering both calcium
and hydroxyl concentrations. Point $D$ corresponds to the
state after infinite hydration time terminating the shown curve.
The progress in time is indicated by arrows. Our numerical results
$(+)$ are in reasonable agreement with the experimental data $(\diamond)$.
We have observed that calculated positions and values for point $C$
depend a) on the employed initial water/cement weight ratio
and b) on the three reaction rate constants $k_1$, $k_2$ and $k_3$
in a rather complex manner.
Contrary to this point $B$ is found to depend only slightly
on the {\em relative} reaction rate constant $k_1/k_2$ over
four orders of magnitude, see Fig.~\ref{fig:fig3}a.
A consequence of the much smaller solubility of $CSH$ as compared to
the second hydrate $CH$ is that in the period $AB$ only {\em two}
chemical reaction are operative. Reaction Eq.~(\ref{ch-precipitation})
is inoperative.
For $k_1/k_2$ ranging between $0.1$ and $10$ the maximum silica
concentrations take constant values of about $1\,mmol/liter$ while
for $k_1/k_2=10^3$ we find $\approx 3\,mmol/liter$.
We have also performed simulations for similar $k_1/k_2$ values employing
different {\em absolute} reaction rate constants, see
Fig.~\ref{fig:fig3}a. We find a
relatively good data collapse over four orders of magnitude.
This result is important insofar as the reaction rate constants have
not been experimentally measured yet. In Fig.~\ref{fig:fig3}b we plot
the momentaneous calcium concentration
versus the silica concentration at point $B$ for various reaction
rate constants. The data are centered around a straight line of slope
$1/3$. Comparison of this result with the stoichiometric coefficients in
Eq.~(\ref{dissolution-reaction}) shows that the silica/calcium
ratio of ions is determined by the dissolution reaction.

\begin{figure}[htb]
\centerline{\psfig{file=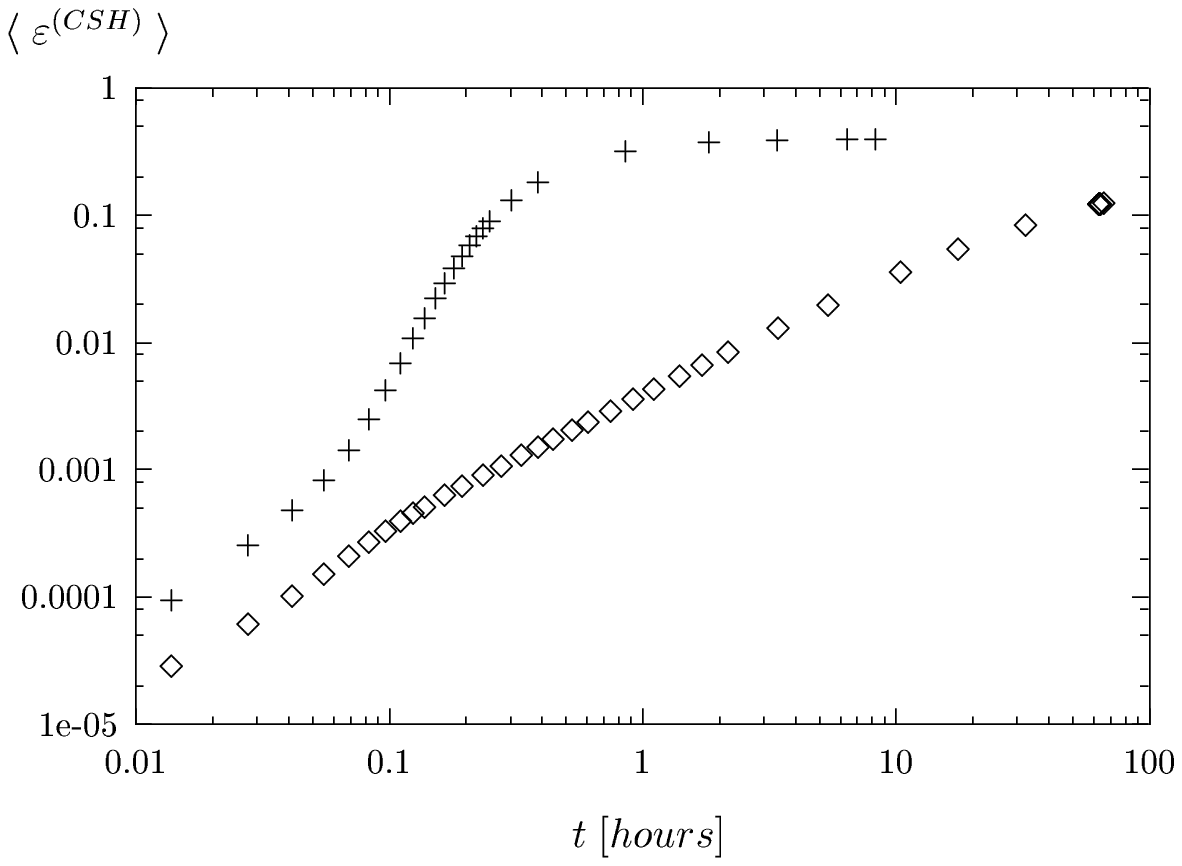,width=\textwidth}}
\caption{
Double logarithmic plot of the average $CSH$ volume
fraction $\langle \;\varepsilon^{(CSH)}\;\rangle$ versus hydration
time $t$ for two sets of reaction rate constants: $(\diamond)$
$k_1=5\cdot 10^{-5}$, $k_2=10^{-6}$ and $k_3=10^{-8}$; $(+)$
$k_1=10^{-2}$, $k_2=10^{-16}$ and $k_3=10^{-13}$; units in
$mol\,m^{-2}\,s^{-1}$. All other parameters as in Fig.~\ref{fig:fig1}.
          }
\label{fig:fig4}
\end{figure}

The characterisation of the
temporal advancement of the hydration process is an important
problem. We show in Fig.~\ref{fig:fig4}
some preliminary results for the time dependence of the mean volume
fraction $\langle \varepsilon^{(CSH)}\rangle$
of the cementious hydrate $CSH$.
For `moderate' reaction constants $(\diamond)$ the hydration is
found to be relatively slow, i.e., after $70\, hours$
of hydration we find only $10\%$ $CSH$.
This value is approximately two times
lower than the experimental one observed by NMR\cite{dobs}.
We note that both
the hydration curve $(\diamond)$ in Fig.~\ref{fig:fig4}
and the parametric curve $(+)$ in Fig.~\ref{fig:fig1}
belong to the same simulation.
Apparently the hydration advancement for `moderate' reaction rates
is nearly constant in time, see curve $(\diamond)$ in Fig.~\ref{fig:fig4}.
For comparison we also show in Fig.~\ref{fig:fig4} a hydration
curve for `fast' hydration $(+)$. After a few minutes
of `induction' period
the hydration rapidly accelerates
and goes already after $1\,hour$ to completion (remaining $C_3S$ less
than $1\;\%$). This process is much to fast and it leads to typical
ion concentrations of $mol/liter$ which are unrealistic high.

As long as the reaction rate constants have not been estimated from
experiments yet one main difficulty in the modellization of cement
hydration consists in finding appropriate values for the rate constants,
{\em which do not contradict} experimental measurements
of mean ion concentrations and hydration advancement. However, the
experimental results belong to hydration in space while
our calculations correspond to two dimensional hydration.
This point is being currently investigated.

\section{Conclusions}\label{conclusions}
We have presented a general, heterogeneous reaction-diffusion model
for solid phase transformation
due to chemical dissolution/precipitation
reactions. The model has focussed on the
important industrial problem of Portland cement hydration though it is
more generally applicable to water/mineral systems.

We have tried to develop an `open' approach based on physical and
chemical considerations (mainly the laws of mass conservation and mass
action). The model includes in its present form
on a coarse grained
length and time scale the full spatial
distribution of solid and liquid phases,
the three main chemical dissolution/precipitation
reactions, and the transport of ions due to diffusion.
The proposed approach naturally allows to include for a varity
of more or less important phenomena depending on imposed
conditions such as solvent flow effects, electrochemical effects,
exothermic effects including heat conduction, pressure effects,
inert filler effects etc..

We have tried to incorporate a reasonable amount of
specific information about the cement hydration into the
investigated model such as
the stoichiometry and kinetics of Portland cement dissolution, the
precipitation/dissolution of the main cementious hydrate (CSH) and
of Portlandite (CH), their approximate solubilities and
molecular volumes, approximate values for the ion
diffusivities in aqueous solution, and initial conditions for water
immersed cement particles (sizes and spatial positions).

The presented results demonstrate considerable richness and complexity
of the cement hydration phenomenon close to controlled
experimental situations.

We have presented some calculated cement micro structures as they
evolve in time. Nucleation and growth of hydrates is found to be
strongly heterogeneous in agreement with experimental observations.
The problem of `autocatalytic' effects on the precipitation processes needs
further investigation.

The presented parametric plots for the average
concentrations of ions in solution
are in qualitative agreement with recent indirect experimental
observations made for stirred cement suspensions\cite{b1}.
The calculated maximum silica concentrations and their
corresponding calcium
concentrations are in reasonable quantitative agreement with experimental
values.
For high dissolution and low precipitation rate constants
we find unacceptable high ion concentrations of order $mol/liter$.
One way to circumvent this problem would be the introduction
of a {\em finite} solubility for $C_3S$ in
Eq.~(\ref{dissolution-kinetics}), however, such a solubility constant
has not been experimentally determined yet.

Furthermore we have investigated the variations of the maximum silica
concentrations for
various reaction rate constants.
The concentrations are found to vary only
slightly with the relative reaction rate constant $k_1/k_2$
over four orders of magnitude.
It would be very interesting to see how the second characteristic
hydration point (point $C$ in Fig.~\ref{fig:fig1}) depend
on the initial water/cement ratio and on the employed reaction
rate constants.

Finally we presented first results on overall hydration curves.
This curves are not very realistic yet. It would be very helpfull
to have experimental order of magnitude estimates for the three
(unknown) reaction rate constants.

For future work we are planning to conduct calculations
in three dimensions with improved kinetic equations.

\section*{Acknowledgments}
We would like to acknowledge stimulating and interesting discussions
with Ch. Vernet, H. Van Damme, S. Schwarzer, A. Nonat and D. Damidot.
F.T. would also like to acknowledge financial support from
CNRS, from GDR project 'Physique des
Milieux H\'et\'erog\`enes Complexes' and from Lafarge Coppee Recherche.
%
%       BIBLIOGRAPHY
%

\end{multicols}
\vfill\eject
\begin{table}\label{table1}
\caption{
Stoichiometric numbers $\nu_1^{(\alpha)}$, $\nu_2^{(\alpha)}$
and $\nu_3^{(\alpha)}$
of the chemical reactions
Eqs.(\ref{dissolution-reaction}), (\ref{csh-precipitation}) and
(\ref{ch-precipitation}) respectively. The last three colums give the
molecular volumes, the diffusion constants and the solubilities at
room temperature and normal pressure.
}
\begin{center}
\begin{tabular}{ccccccc}
Species   & &  Stoichiometric numbers & &  Molecular volume &
Diffusion constant & Solubility \\
$\alpha$  &$\nu_1^{(\alpha)}$&$\nu_2^{(\alpha)}$&$\nu_3^{(\alpha)}$&
[$10^{-3}\;liter/mol$] & [$10^{-10}\;m^2/s$]&[$10^{-3}\;
mol/liter$] \\ \hline
$H_2O$    & $-3$ & $-1$ & $0$ &    $18.01$ & $   0 $  &  ---    \\
$C_3S$    & $-1$ & $0$  & $0$ &    $73.1^{~(b)}$ & $   0 $  &
$\infty^{~(f)}$ \\
$C_{1.5}SH_{2.5}$     & $0$  & $1$& $0$ &   $80.40^{~(c)}$ & $   0 $
&  $2.4^{~(d)}$    \\
$CH$      & $0$  & $0$  & $1$ &$33.07^{~(a)}$ & $   0 $  &  $20.2^{~(a)}$   \\
$\ca$     & $3$  &$-3/2$& $-1$&      ---  & $  7.9^{~(e)}$  &   ---     \\
$\oh$     & $4$  & $-1$ & $-2$&      ---  & $ 53.0^{~(e)}$  &   ---     \\
$\si$     & $1$  & $-1$ & $0$ &      ---  & $ 5.0^{~(f)}$  &   ---    \
\end{tabular}
\end{center}
\end{table}
\small{
\noindent
$^{~(a)}$ H.F.W. Taylor, {\em Cement Chemistry}, (Academic Press,
London, 1990), pp.~125;
$^{~(b)}$ {\em ibid}, p.~15;
$^{~(c)}$ {\em ibid}, p.~152, we assume for the calculations a density of
$2.35\,g\,cm^{-3}$;
$^{~(d)}$ We have calculated this value from the supersaturation
concentrations of calcium and silicate ions, as reported
in Ref.~\ref{b8};
$^{~(e)}$ R.A. Alberty and R.J. Silbey, {\em Physical
    Chemistry}, (Wiley\&Sons, New York, 1992), pp.~839;
$^{~(f)}$ Measured values are not known to us.
        }
\vfill\eject

%
%       FIGURE CAPTIONS
%
\vfill\eject

\vfill\eject

\vfill\eject

\vfill\eject


\begin{thebibliography}{10}
\bibitem{b13}\label{b13}
  As aforementioned we determine the solvent volume and its mole
  numbers exclusively from Eq.(\ref{solvent-volume}). This assumption
  is reasonable as long as the number of solvated ions is much lower
  than the number of solvent molecules.
\bibitem{b17}\label{b17}
  It has been reported that under very particular experimental
  conditions (strongly diluted suspensions)
  the precipitation of  calcium hydroxide becomes supressed,
  see Ref.~\ref{b9}.
\end{thebibliography}
\end{document}